%% file: paper.tex
\documentstyle[epsfig,subfig,amsmath]{elsarticle}
\newcommand{\newc}{\newcommand}
\newc{\lra}{\leftrightarrow}
\newc{\beq}{\begin{equation}}
\newc{\eeq}{\end{equation}}
\newc{\barr}{\begin{eqnarray}}
\newc{\earr}{\end{eqnarray}}

\input{begin.tex}

\begin{document}
\def\vbf{\mbox{\boldmath $\upsilon$}}
\def\barr{\begin{eqnarray}}
\def\earr{\end{eqnarray}}
\def\g{\gamma}
\newcommand{\dphi}{\delta \phi}
\newcommand{\bupsilon}{\mbox{\boldmath \upsilon}}
\newcommand{\bfup}{\mbox{\boldmath \upsilon}}
\newcommand{\at}{\tilde{\alpha}}
\newcommand{\pt}{\tilde{p}}
\newcommand{\Ut}{\tilde{U}}
\newcommand{\rhb}{\bar{\rho}}
\newcommand{\pb}{\bar{p}}
\newcommand{\pbb}{\bar{\rm p}}
\newcommand{\kt}{\tilde{k}}
\newcommand{\wt}{\tilde{w}}

\date{\today}
\title {Modulation and diurnal variation  in axionic dark matter searches}
%
%
%
%
%
\author{Y. Semertzidis$^{1} and J.D. Vergados${1,2}}
%
%
\address{
$^1$ Center for Axion and Precision Physics Research, IBS, Daejeon 305-701, Republic of Korea
}
\address{$^2$ Centre for the Subatomic Structure of Matter (CSSM), University of Adelaide, Adelaide SA 5005, Australia\footnote{
Theoretical Physics,University of Ioannina, Ioannina, Gr 451 10, Greece.
\\E-mail:Vergados@uoi.gr}}
\begin{frontmatter}
\begin{abstract}
In the present work we study possible time dependent effects in Axion Dark Matter searches  employing  resonant cavities. We find that the width of the resonance, which depends on the axion mean square  velocity in the local frame, will show an annual variation due to the motion of the Earth  around the sun (modulation). Furthermore, if the experiments become  directional, employing suitable resonant cavities, one expects large asymmetries in the observed widths relative to the sun's direction of motion. Due to the rotation of the Earth around its axis, these asymmetries will manifest themselves as a diurnal variation in the observed width.
\end{abstract}

\end{frontmatter}

\section{Introduction}
In the standard model there is a source of CP violation from the phase in the Kobayashi-Maskawa mixing matrix. This, however, is not large enough to explain the baryon asymmetry observed in nature.
Another source is the phase in the interaction between gluons ($\theta$-parameter), expected to be of order unity.
The non observation of elementary electron dipole moment limits its value to be $\theta\le≤10^{-9}$ this has been known as the strong CP problem. 
A solution to this problem has been the P-Q (Peccei-Quinn) mechanism. 
In extensions of the S-M, e.g. two Higgs doublets, the Lagrangian has a global P-Q chiral symmetry $U_{PQ}(1)$, which is spontaneously broken, generating a Goldstone boson, the axion (a). In fact 
the axion has been proposed a long time ago as a solution to the strong CP problem \cite{PecQui77} resulting to a pseudo Goldstone Boson \cite{SWeinberg78,Wilczek78,PWW83,AbSik83,DineFisc83}. \\
QCD effects violate the P-Q symmetry and generate a potential $(m_a a)^2 /2 $ for the axion field $a=\theta f_a$ with  axion mass $ma =(\Lambda^2_{QCD})/f_a $with minimum at $\theta=0$. Axions can be viable if the SSB (spontaneous symmetry  breaking) scale is large $f_a \geq 100$ GeV.
Thus the  axion becomes a pseudo-Goldstone boson
An initial displacement $a_i=\theta _if_a$  of the axion field causes an oscillation with frequency $\omega= ma $ and energy density $ρ\rho_D = (\theta if_am_a)^2 /2 $
The production mechanism varies depending on when SSB takes place, in particular whether it takes before or after inflation.

\begin{figure}
\begin{center}
\includegraphics[width=1.0\textwidth, height=0.4\textwidth]{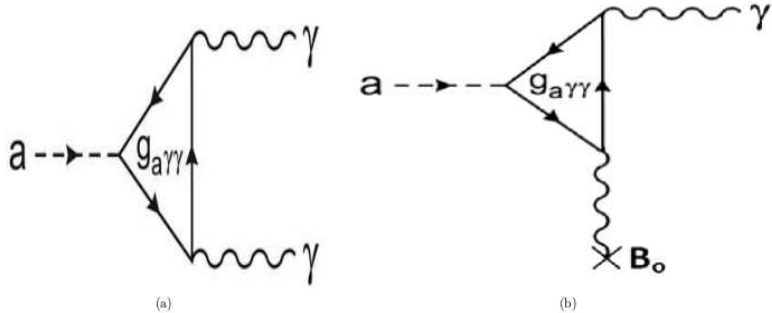}
 \caption{The  axion decay to photons, axion photon coupling, (a) and the  axion to photon conversion  in the presence of a magnetic field, the Primakoff effect, (b).}
 \label{AxionPhoton}
  \end{center}
  \end{figure}

 The axion field is homogeneous over a large de Broglie wavelength, oscillating in a coherent way, which makes it  ideal cold dark matter candidate  in the mass range $10^{-6}$ eV$\leq m_a \leq 10^{-3}$ eV. 
In fact it has been recognized long time ago  as a prominent  dark matter candidate  by Sikivie \cite{Sikivie83}, and others, see, e.g, \cite{PriSecSad88}.
The axions are extremely light. So it is impossible to detect dark matter axions via scattering them off targets. 
They are detected by their conversion to photons in the presence of a magnetic field (Primakoff effect), see Fig. \ref{AxionPhoton}. The produced photons are detected in a resonance cavity  as suggested by Sikivie \cite{Sikivie83}.

\begin{figure}[!ht]
\begin{center}
\includegraphics[width=\textwidth, height=0.5\textwidth]{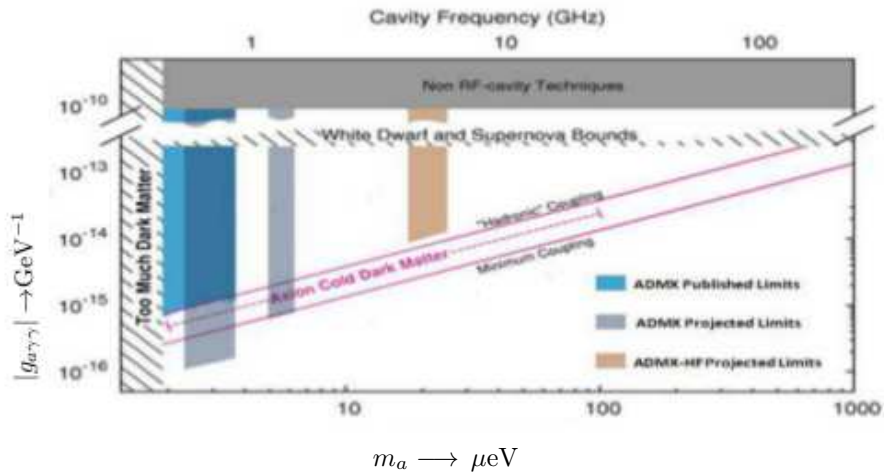}
{\hspace{-2.0cm} {$m_{a}\longrightarrow\,\mu$eV}}\\
 \caption{The  parameter space relevant for axion  as dark matter candidate.}
 \label{fig:AxionParSp}
  \end{center}
  \end{figure}

 In fact various experiments\footnote{Heavier axions with larger mass in the 1eV region produced thermally ( such as via the  $a\pi\pi\pi$ mechanism), e.g. in the sun, are also interesting and are searched by CERN Axion Solar Telescope (CAST) \cite{CAST11}.
Other axion like particles (ALPs), with broken symmetries not connected to QCD, and dark photons form  dark matter candidates called WISPs (Weakly Interacting Slim Particles) , see, e.g.,\cite{ADDKM-R10}, are also being searched.}
 such as ADMX and ADMX-HF collaborations 
\cite{Stern14,MultIBSExp}, \cite{ExpSetUp11b},\cite{ADMX10} are now planned to search for them. In addition, the newly established center for axion and physics research (CAPP)  has started an ambitious axion dark matter research program \cite{CAPP}, using SQUID and HFET technologies \cite{ExpSetUp11a}.
  The allowed parameter space has been presented in a nice slide by Raffelt \cite{MultIBSTh} in the recent Multidark-IBS workshop and, focusing on the axion as dark matter candidate, by Stern \cite{Stern14} (see Fig.  \ref{fig:AxionParSp}, derived from Fig. 3 of ref \cite{Stern14}).
	  \begin{figure}[!ht]
\begin{center}
\includegraphics[width=\textwidth, height=0.5 \textwidth]{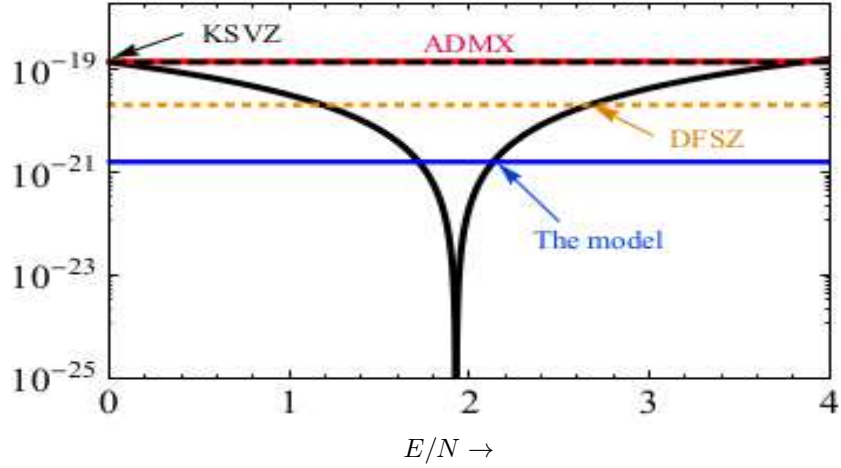}
{\hspace{-2.0cm} {$E/N\rightarrow$}}\\
 \caption{The  ratio $\frac{g_{a\gamma\gamma}^2}{m^2_a}$, in units ofGeV $^{-2}/$eV$^2$,  as a function of $\frac{E}{N}$, where $E$ is the axion electromagnetic anomaly and $N$ is the color anomaly number.}
 \label{fig:gaxion}
  \end{center}
  \end{figure}
		  \begin{figure}[!ht]
\begin{center}
\includegraphics[width=\textwidth, height=0.5 \textwidth]{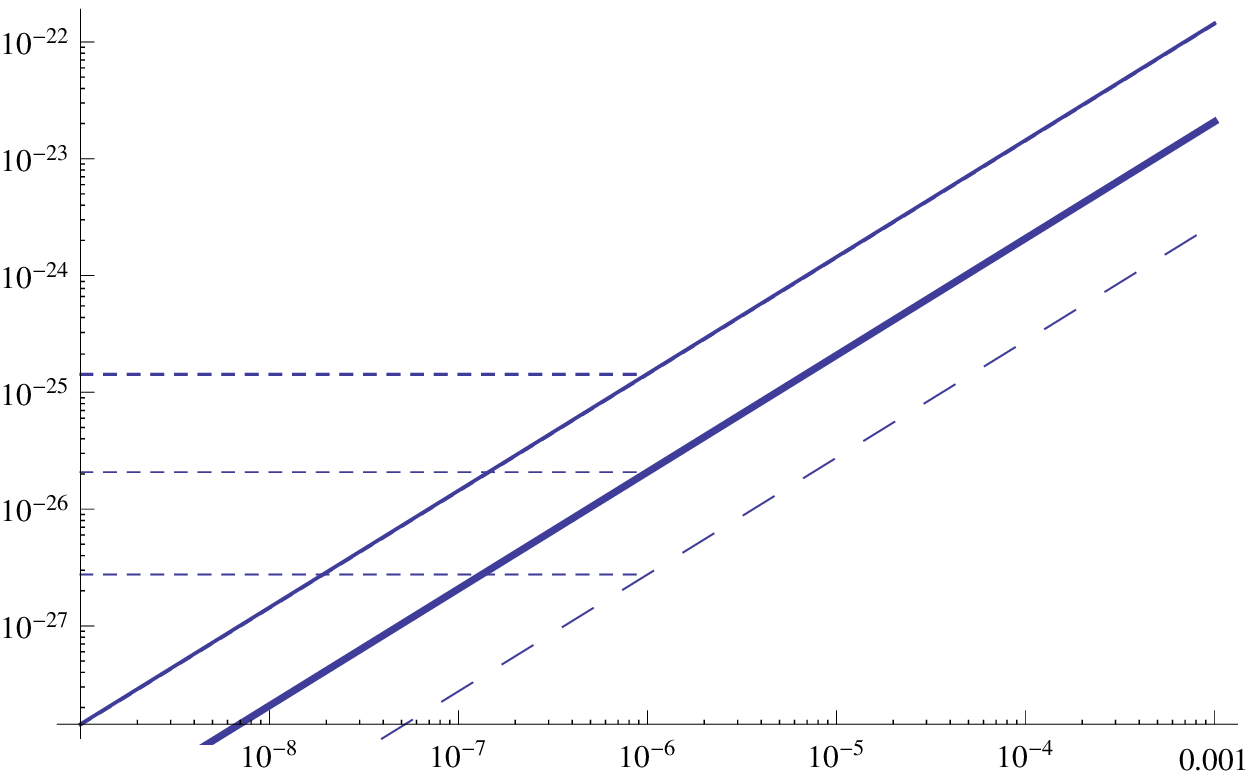}
{\hspace{-2.0cm} {$m_{a}\longrightarrow$ eV}}\\
 \caption{The  ratio $\frac{g_{a\gamma\gamma}^2}{m_a}$ in units of GeV$^{-2}/$eV as a function of $m_a$. The curves correspond to the current experimental limit (solid line) and three theoretical predictions extracted from Fig. 3  of ref \cite{Ahn14}. The solid, thick solid and dashed curve correspond to E/N=0,\,8/3 and 112/51 respectively. The first almost coincides with that extracted from experiment, while the last one is proposed as theoretically favored. It is seen that for $m_a=10^{-6}$ eV the power produced can drop by almost two orders of magnitude below the current experimental limit, if the theoretically favored curve (dashed line).}
 \label{fig:gsqaxion}
  \end{center}
  \end{figure}
In the present work we will take the view that the axion is non relativistic with mass in $\mu$eV-meV scale moving with an average velocity which is $\approx0.8\times10^{-3}$c.  The width of the observed resonance depends on the axion mean square velocity in the local frame. Thus one expects it to exhibit a time variation due to the motion of the Earth. Furthermore in directional experiments involving long cavities, one expects asymmetries with regard to the sun's direction of motion as it goes around the center of the galaxy. Due to the rotation of the Earth around its axis these asymmetries in the width of the resonance will manifest themselves in their diurnal variation. These two special signatures, expected to be sizable,  may aid the analysis of axion dark searches in discriminating against possible backgrounds.
\section {Brief summary of the formalism}
The photon axion interaction is dictated by the Lagrangian:
\begin{equation}
{\cal L}_{a \gamma\gamma}=g_{a\gamma\gamma}a {\bf E}\cdot{\bf B},\, g_{a\gamma\gamma}=\frac{\alpha g_{\gamma}}{ \pi f_a},
\end{equation}
where  ${\bf E}$ and ${\bf B}$ are the electric and magnetic fields,  $f_a$ the axion decay constant and $g_{\gamma}$ a model dependent constant of order one 
\cite{Stern14},\cite{HKNS14},\cite{JEKim98},\cite{RosBib00} given by:
\begin{equation}
g_{\gamma}=\frac{1}{2}\left(\frac{E}{N}-\frac{2}{3}\frac{4+z}{1+z} \right )
\end{equation}
where $z$ is the ratio of the up and down quark masses, $N$ is the color axion anomaly and $N$ is the axion EM anomaly. The second $z$-dependent term \cite{RosBib00} is 1.95. In grand unifiable models, like the DFZX axion, $E/N$  is 8/3. In the case of the KSFZ axion $E/N=0$, while in the flavor Peccei-Quin (flavored PQ) symmetry model to be discussed below,  $E/N=112/51$. We thus see that the $g^2_{\gamma}$ range is given by:
\begin{equation}
g^2_{\gamma}=\left \{ \begin{array}{rl} \mbox{DFSZ:}&0.13 \\
\mbox{KSVZ:}&0.94\\ \mbox{Flavored PQ:}&0.061\\ \end{array} \right .
\label{Eq:ggammasq}
\end{equation}
 Axion dark matter detectors \cite{HKNS14} employ an external magnetic field, ${\bf B}\rightarrow {\bf B}_0$ in the previous equation, in which case one of the photons is replaced by a virtual photon, while the other maintains the energy of the axion, which is its mass plus a small fraction of kinetic energy. \\
The power produced, see e.g.  \cite{Stern14}, is given by:
\begin{equation}
P_{mnp}=g_{a\gamma\gamma}^2\frac{\rho_a}{m_a} B_0^2 V C_{mnp} Q_L
\label{Eq:Pmnp}
\end{equation}
$Q_L$ is the loaded quality factor of the cavity. Here we have assumed $Q_L$ is smaller than the axion width $Q_a$, see below.  More generally, $Q_L$ should be substituted by min ($Q_L$, $Q_a$).
This power depends on the axion particle density $n_a={\rho_a}/{m_a}$ with density $\rho_a$ assumed to be the same with that used for WIMPs (Weakly Interacting Massive Particles), inferred from the rotational curves, i.e. the axion particle density is much larger than that expected for WIMPs. In any case the power produced  is pretty much independent of the velocity and the velocity distribution.

 In addition to the axion density in our position, which can somehow  be determined from the rotation curves, it is a function of the theoretical parameter $ {g_{a\gamma\gamma}^2}/{m_a} $. The study of this parameter has been the subject of many theoretical studies (see e.g. the recent works \cite{HKNS14},\cite{JEKim98} and references there in). Recently, however, an unconventional but  economical extension of the Standard Model of particle physics has been proposed \cite{Ahn14}, which attempts  to  deal with the fermion mass hierarchy problem and the strong CP problem, in a way that no domain wall problem arises, in a supersymmetric framework involving the $A_4\times U(1)_X$ symmetry. The global $ U(1)_X$ symmetry, which can tie the above together is referred to as flavored Peccei-Quinn (Flvored PQ) symmetry.  In this model the flavon fields, which are responsible for the spontaneous symmetry breaking of quarks and leptons, charged under $U(1)_X$, connect the various expectation values. Thus the Peccei-Quinn symmetry is estimated to be located around $10^{12}$ GeV through its connection to the fermion masses obtained in the context of $A_4$. They are thus able to obtain the ratio of $ {g_{a\gamma\gamma}^2}/{m^2_a} $, as a function of E/N (see Eq. (\ref{Eq:ggammasq}) and  Fig. \ref{fig:gaxion}, based on Fig. 3 of  Flavored PQ       symmetry \cite{Ahn14} ).   From this figure we can extract  a relationship between    $ {g_{a\gamma\gamma}^2}/{m_a} $ and $m_a$. This is shown in Fig.   \ref {fig:gsqaxion}. From this figure we see that the power produced may be almost two orders of magnitude below the current experimental limit.

Admittedly this $A_4\times U(1)_X$ symmetry is introduced ad hoc and especially the introduction $U(1)_X$ is not particularly otherwise motivated. Furthermore the fermion masses and paricularly the neutrino mass are not fully understood. Anyway the experiments may have to live with such a pessimistic prospect and we may have to explore other signatures of the process. The power spectrum comes to mind.                                                                                         .
	
The axion power spectrum, which is of great interest to experiments, is written as a Breit-Wigner shape \cite{HKNS14}, \cite{KMWM85}:
\begin{equation}
\left |{\cal A}(\omega) \right |^2=\frac{ \rho_D}{m_a^2}\frac{\Gamma}{(\omega-\omega_a)^2+(\Gamma/2)^2},\Gamma=\frac{\omega_a}{Q_a}
\end{equation}
with $\omega_a=m_a\left (1+(1/6)\prec\upsilon^2\succ\right )\approx m_a$ giving the location of the maximum of the spectrum and $Q_{a}=1/(\prec\upsilon^2\succ/3),\,\Gamma=m_a (\prec\upsilon^2\succ/3).$\\
Since  in the
axion DM search case the cavity detectors have reached such a very high energy resolution \cite{Duffy05,Duffy06}, one should try to accurately evaluate the width of the expected power spectra in various theoretical models.\\
The width explicitly depends on the average axion velocity squared,  $\prec \upsilon^2\succ$. With respect to the  galactic frame $\prec \upsilon^2\succ$ takes the usual value of $(3/2 )\upsilon_0^2  $. In the laboratory frame, taking the sun's motion  into account, we find  $\prec \upsilon^2\succ =(5/2 )\upsilon_0^2$, i.e. the width in the laboratory is affected by  the sun's motion.
If we take into account the motion of the Earth around the sun $\prec \upsilon^2\succ$ depends on the phase of the Earth and, correspondingly, the width becomes time dependent (modulation) as described below (see section  \ref{sec:mod}).\\
 The situation becomes more dramatic as soon the experiment is directional. In this case the width depends strongly on the direction of observation relative to the sun's direction of motion.
 Directional experiments can, in principle, be performed by changing the orientation of a long  cavity \cite{Stern14},\cite{FutExp14},\cite{IrasGarcia12}, provided that the axion wavelength is not larger than the length of the cylinder, $\lambda_a\le h$. In the ADMX \cite{Stern14} experiment $h=100$cm, while  from their Fig. 3 one can see that  the relevant for dark matter wavelengths $\lambda_a$ are  between 1 and 65 cm.
\section{Modification of the width due to the motion of the Earth and the sun.}
From the above discussion it appears that velocity distribution of axions may play a role in the experiments. 
\subsection{The velocity distribution}
If the axion is going to be considered as dark matter candidate, its density should fit the rotational curves.  Thus for temperatures $T$  such that $m_a/T\approx 4 \times 10^{6}$ the velocity distribution can be taken  to be   analogous to that assumed for WIMPs. In the present case we will consider only   a M-B distribution in the galactic frame with a characteristic velocity which equals the velocity of the sun around the center of the galaxy, i.e. $\upsilon_0 \approx 220$km/s.  Other possibilities such as, e.g., completely phase-mixed DM, dubbed ``debris flow''
(Kuhlen et al.~\cite{Spergel12}), and caustic rings  (Sikivie~\cite{SIKIVI1},\cite{SIKIVI2}\cite{Sikivie08}, Vergados~\cite{Verg01} etc are currently under study and hey will appear elsewhere\cite{VerSemSik}. So we will employ here the distribution:
\begin{equation}
f({\vec \upsilon})=\frac{1}{(\sqrt{\pi}\upsilon_0)^3}e^{-\frac{\upsilon^2}{\upsilon^2_0}}
\end{equation}
In order to compute the average of the velocity squared entering the power spectrum we need to find the local velocity distribution by taking into account the velocity of the Earth around the sun and the velocity of the sun around the center of the galaxy. The first motion leads to a time dependence of the observed signal in standard experiments , while the latter motion leads to asymmetries in directional experiments. 

\subsection{The annual modulation in non directional experiments}
\label{sec:mod}
The modification of the velocity distribution in the local frame due to annual motion of the Earth is expected to affect the detection of axions in a time dependent way, which, following the terminology of the standard WIMPs, will be called the modulation effect 
\cite{DFS86} (the corresponding effect due to the rotation of the Earth around its own axis is too small to be observed). Periodic signatures  for the detection of cosmic axions were first considered by Turner \cite{Turner90}.
 
\begin{figure}[!ht]
\begin{center}
\rotatebox{90}{\hspace{-0.0cm} {$y^2 f(y)\longrightarrow$}}
\includegraphics[width=0.9\textwidth,height=0.3\textwidth]{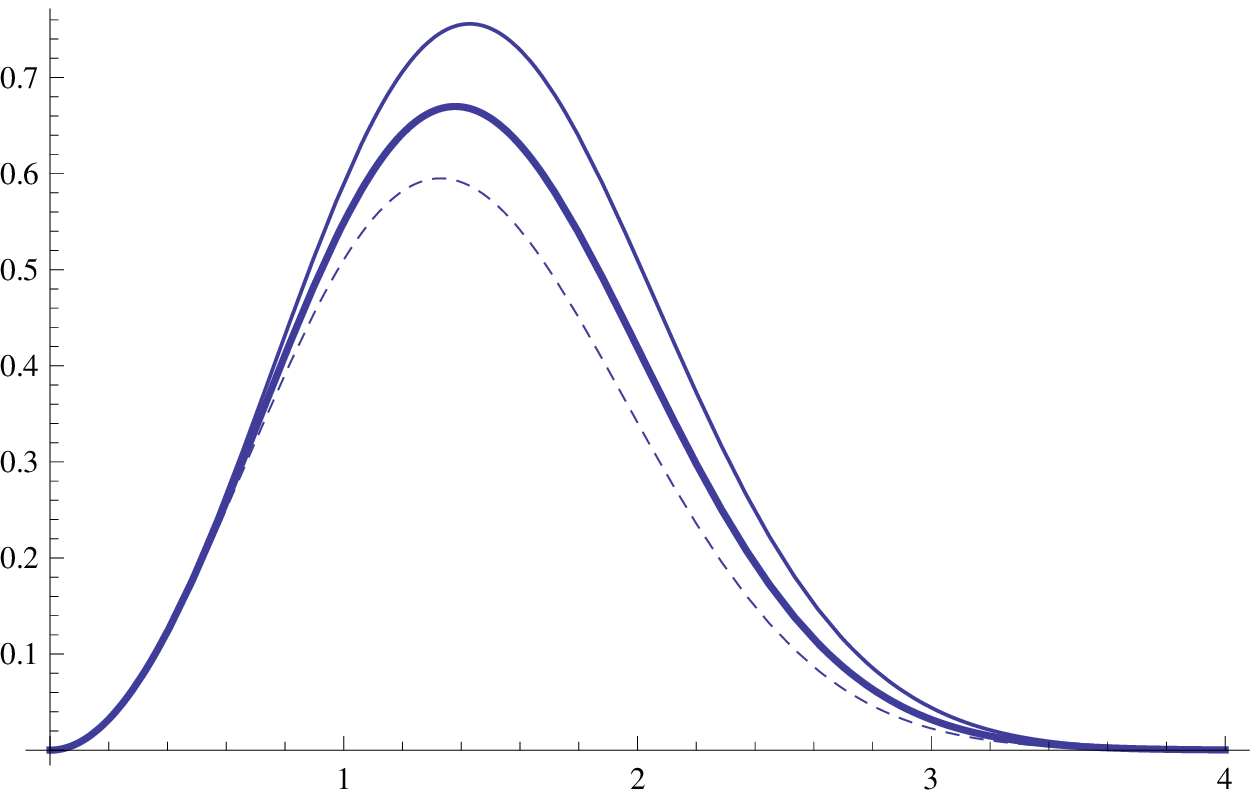}
\\
{\hspace{-2.0cm} {$y=\upsilon/\upsilon_0\longrightarrow$ }}\\
 \caption{The axion velocity distribution in the local frame. It is changing with time  due to the motion of the Earth and the sun. We exhibit  here the distribution in June  (solid line), in December  (thick solid line) and in September or March (dotted line). The last curve coincides with that in which the motion of the Earth is ignored. 
}
\label{fig:modvel}
  \end{center}
  \end{figure}

\begin{figure}[!ht]
\begin{center}
\rotatebox{90}{\hspace{-0.0cm} {$\Gamma(\alpha)/\prec\Gamma\succ\longrightarrow$}}
\includegraphics[width=0.9\textwidth,height=0.5\textwidth]{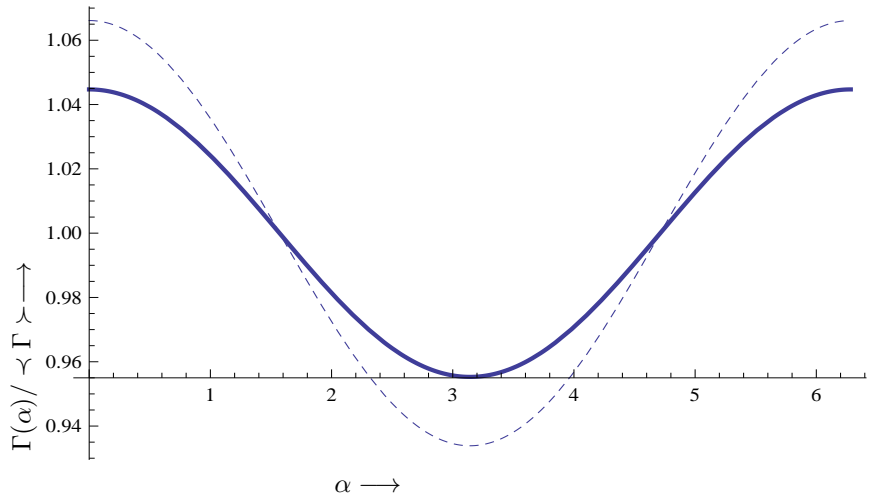}
\\
{\hspace{-2.0cm} {$\alpha\longrightarrow$ }}\\
 \caption{The ratio of the modulated width divided by the time averaged width as a function of the phase of the Earth. The solid line corresponds to the standaed M-B distribution and the dotted line to an axially symmetric M-B distribution with asymmetry parameter $\beta=0.5$ (see text).}
\label{fig:modRW2}
  \end{center}
  \end{figure}

So our next task is to transform the velocity distribution from the
galactic to the local frame. The needed equation, see e.g.
\cite{Vergados12}, is:
  \begin{equation}
{\bf y} \rightarrow {\bf y}+{\hat\upsilon}_s+\delta \left
(\sin{\alpha}{\hat x}-\cos{\alpha}\cos{\gamma}{\hat
y}+\cos{\alpha}\sin{\gamma} {\hat \upsilon}_s\right ) ,\quad
y=\frac{\upsilon}{\upsilon_0} \label{Eq:vlocal} 
\end{equation} 
with
$\gamma\approx \pi/6$, $ {\hat \upsilon}_s$ a unit vector in the
Sun's direction of motion, $\hat{x}$  a unit vector radially out
of the galaxy in our position and  $\hat{y}={\hat
\upsilon}_s\times \hat{x}$. The last term in the first expression
of Eq. (\ref{Eq:vlocal}) corresponds to the motion of the Earth
around the Sun with $\delta$ being the ratio of the modulus of the
Earth's velocity around the Sun divided by the Sun's velocity
around the center of the Galaxy, i.e.  $\upsilon_0\approx 220$km/s
and $\delta\approx0.135$. The above formula assumes that the
motion  of both the Sun around the Galaxy and of the Earth around
the Sun are uniformly circular. The exact orbits are, of course,
more complicated  but such deviations are not
expected to significantly modify our results. In Eq.
(\ref{Eq:vlocal}) $\alpha$ is the phase of the Earth ($\alpha=0$
around the beginning of June)\footnote{One could, of course, make the time
dependence of the rates due to the motion of the Earth more
explicit by writing $\alpha \approx(6/5)\pi\left (2 (t/T)-1 \right
)$, where $t/T$ is the fraction of the year.}.
 \\The velocity distribution in the local frame is affected by the motion of the Earth as exhibited in Fig. \ref{fig:modvel} at four characteristic periods.  
	Appropriate in the analysis of the experiments is the relative  modulated width, i.e. the ratio of the time dependent width divided by the time averaged with, is  shown in Fig. \ref{fig:modRW2}. The results shown here are for spherically symmetric M-B distribution as well as an axially symmetric one with asymmetry parameter $\beta=0.5$ with
	\begin{equation}
 \beta=1-\frac{\langle \upsilon_t^2\rangle}{2 \langle \upsilon_r^2\rangle}
 \end{equation}
with  $\upsilon_r$ the  radial, i.e. radially out of the galaxy, and $\upsilon_t$ the tangential  component of the velocity.
Essentially similar results are obtained by more exotic models, like a combination of M-B and Debris flows considered by Spegel and collaborators \cite{Spergel12}. We see that the effect is  small, around 15$\%$ difference between maximum and minimum in the presence of the asymmetry,  but still larger than that expected in ordinary dark matter searches. If we do detect the axion frequency, then we can determine its width with high accuracy and detect its modulation as a function of time.
\subsection{Asymmetry of the rates in directional experiments}
\begin{figure}[!ht]
\begin{center}
\rotatebox{90}{\hspace{-0.0cm} {$r(\Theta,\Phi)=\Gamma_dir/\Gamma_{st}\rightarrow$}}
\includegraphics[width=0.9\textwidth,height=0.5\textwidth]{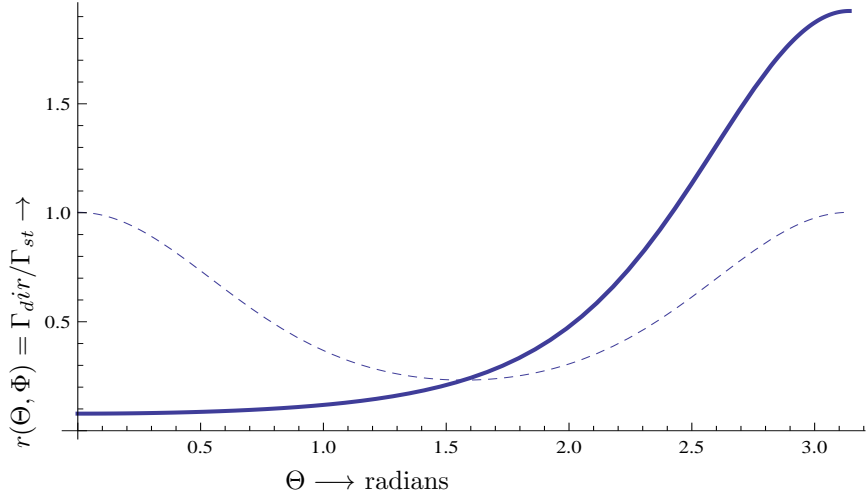}\\
{\hspace{-2.0cm} {$\Theta\longrightarrow$ radians}}\\
 \caption{The  ratio of the width expected in a directional experiment divided by that expected in a standard experiment. The solid line is expected, if the sense of direction is known, while the dotted will show up, if the sense of direction is not known.}
 \label{fig:Axionwidth}
  \end{center}
  \end{figure}

Consideration of the velocity distribution will give an important signature, if directional experiments become feasible. This can be seen as follows:
\begin{itemize}
\item The width will depend  specified by two angles $\Theta$ and 
$\Phi$. \\
The angle $\Theta$  is the polar angle  between the sun's velocity and the direction of observation. The angle $\Phi$ is measured in a plane perpendicular to the sun's velocity, starting from the line coming radially out of the galaxy and passing through the sun's location.
\item The axion velocity, in units of the solar velocity, is given as \\
\begin{equation}
{\bf y}=y\left ({\hat x}\sqrt{1-\xi^2}\cos{\phi}+ {\hat y}\sqrt{1-\xi^2}\sin{\phi}+{\hat z}\xi \right  )
\end{equation}
\item $\mbox{Ignore the motion  of the Earth around the sun, i.e. }\delta=0$
Then the velocity distribution in the local frame is obtained by the substitution:\\
\begin{equation}
\upsilon^2 \rightarrow\upsilon_0^2\left(y^2 +1+2y\left (\xi \cos{\Theta}+\sqrt{1-\xi^2}\sin{\Theta}\left (\cos{\Phi}\cos{\phi}
 +\sin{\Phi}\sin{\phi}\right ) \right )\right )
\end{equation}
One then can integrate over  $\xi$ and $\phi$. The results become essentially independent of $\Phi$, so long as the motion of the Earth around the sun is ignored\footnote{The annual modulation of the expected results  due to the motion of the Earth around the sun will show up  in  the directional experiments as well, but it is going to be less important and it will not be discussed here.}.Thus we  obtain $\prec \upsilon^2\succ$ from the  axion velocity distribution for various polar angles $\Theta$. 
\end{itemize}
We write the width observed in a directional experiment as:
\begin{equation}
\Gamma=r(\Theta)\Gamma_{st}
\end{equation}
where $\Gamma_{st}$ is the width in the standard experiments. Ignoring the motion of the Earth around the sun the factor $r$ depends only on $\Theta$. Furthermore, if for simplicity we ignore the upper velocity bound (cut off)  in the M-B distribution, i.e. the escape velocity $\upsilon_{esc}=2.84 \upsilon_0$, we can get the solution in analytic form. We find:
\begin{equation}
r(\Theta)=\frac{2}{5}\frac {e^{-1}}{2}\left ( e^{\cos ^2{\Theta }} (\cos {2 \Theta }+4) \mbox{ erfc}(\cos \Theta
   )-\frac{2 \cos \Theta }{\sqrt{\pi }}\right ),\, \mbox{(sense  known)},
\end{equation}
with $\mbox{erfc}(z)=1-\mbox{erf}(z), \mbox{erf}(z)=\int_0^z dt e^{-t^2} \mbox{( error function)}$
\begin{equation}
r(\Theta)=\frac{2}{5}\frac{1}{2} e^{-\sin ^2\Theta } (\cos{ 2 \Theta }+4),\, \mbox{(sense of direction not known)},
\end{equation}
The adoption of an upper cut off has little effect. In Fig. \ref{fig:Axionwidth} we present the exact results.
	 The above results were obtained with a M-B velocity distribution\footnote{Evaluation  of the relevant average velocity squared  in  some other  models \cite{Sikivie11},\cite{GHP-W14}, which lead to caustic ring distributions, can also be worked out for axions as above in a fashion
 analogous to that of WIMPs  \cite{Verg01}, but this is not the subject of the present paper}. 

Our  results  indicate that the width will exhibit diurnal variation! For a cylinder of Length $L$ such a variation is expected to be favored \cite{IrasGarcia12} in the regime of $m_a L=10-25\times 10^{-4}$eV-m.
	This diurnal variation will be discussed in the next section.
	
	\section{The diurnal variation in directional experiments}
The apparatus will be oriented in a direction specified in the local frame, e.g. by a point in the sky specified, in the equatorial system, by right ascension $\tilde{\alpha}$ and inclination $\tilde{\delta}$\footnote{We have chosen to adopt the notation $\tilde{\alpha}$ and $\tilde{\delta}$ instead of the standard notation $\alpha$ and $\delta$ employed by the astronomers to avoid possible confusion stemming from the fact that $\alpha$  is usually  used to designate the phase of the Earth and $\delta$  for the ratio of the rotational velocity of the Earth around the Sun  by the velocity of the sun around the center of the galaxy}. This will lead to a diurnal variation\footnote{This should not be confused with the diurnal variation expected even in non directional experiments due to the rotational velocity of the Earth, which is expected to be too small.} of the event rate \cite{CYGNUS09}. This situation has already been discussed in the case of standard WIMPs \cite{VerMou11,VerSem15}. We will briefly discuss the transformation into the relevant astronomical coordinates here.

A vector oriented by $(\tilde{\alpha},\tilde{\delta}) $ in the laboratory  is given   in the galactic frame by a unit vector with components:
\begin{equation}
\left (
\begin{array}{l}
 y \\
 x \\
 z
\end{array}
\right )
=\left (\begin{array}{l}
 -0.868 \cos {\tilde{\alpha} } \cos {\tilde{\delta} }-0.198 \sin {\tilde{\alpha}
   } \cos {\tilde{\delta }}+0.456 \sin{\tilde{\delta} } \\
 0.055 \cos {\tilde{\alpha}} \cos {\tilde{\delta}}+0.873 \sin {\tilde{\alpha}
   } \cos {\tilde{\delta }}+0.4831 \sin {\tilde{\delta }} \\
 0.494 \cos {\tilde{\alpha} } \cos {\tilde{\delta} }-0.445 \sin {\tilde{\alpha}
   } \cos {\tilde{\delta}}+0.747 \sin {\tilde{\delta} }
\end{array}
\right ).
\end{equation}
where $\tilde{\alpha}$ is the right ascension and $\tilde{\delta}$ the inclination.
\\Due to the Earth's rotation the unit vector $(x,y,z)$, with a suitable choice of the initial time, $\tilde{\alpha}-\tilde{\alpha}_0=2\pi(t/T)$, is changing as a function of time

\begin{equation}
x=\cos {\gamma } \cos {\tilde{\delta} } \cos \left(\frac{2 \pi  t}{T}\right)-\sin
   {\gamma } \left(\frac{}{}\cos {\delta } \cos{\theta _P} \sin
   \left(\frac{2 \pi  t}{T}\right)
	+ \sin {\tilde{\delta} } \sin{\theta
   _P}\right),
\end{equation}
\begin{equation}
y=\cos \left(\theta _P\right) \sin {\tilde{\delta }}-\cos {\tilde{\delta} } \sin
   \left(\frac{2 \pi  t}{T}\right) \sin{\theta _P},
\end{equation}
\begin{equation}
z=\cos \left(\frac{2 \pi  t}{T}\right) \cos {\tilde{\delta} } \sin {\gamma }
+\cos
   {\gamma } \left(\cos {\tilde{\delta} } \cos{\theta _P} \sin
   \left(\frac{2 \pi  t}{T}\right)
	+ \sin {\tilde{\delta} } \sin{\theta
   _P}\right ),
\end{equation}
where  $T$ is the period of the Earth's rotation,  $\gamma\approx 33^0$ was given above and $\theta_P=62.6^0$ is the angle the Earth's north pole forms with the axis of the galaxy.
 Thus the angles $\Theta$, which is of interest to us  in directional experiments, is given by
\begin{equation}
\Theta=\cos^{-1}{z},
\end{equation}
An analogous, albeit a bit more complicated, expression dependent on $x,\,y,\,z$ can be derived for the angle  $\Phi$.

\begin{figure}[!ht]
 \begin{center}
\rotatebox{90}{\hspace{-0.0cm} {$\Theta  \longrightarrow$ radians}}
     \includegraphics[width=0.9\textwidth,height=0.5\textwidth]{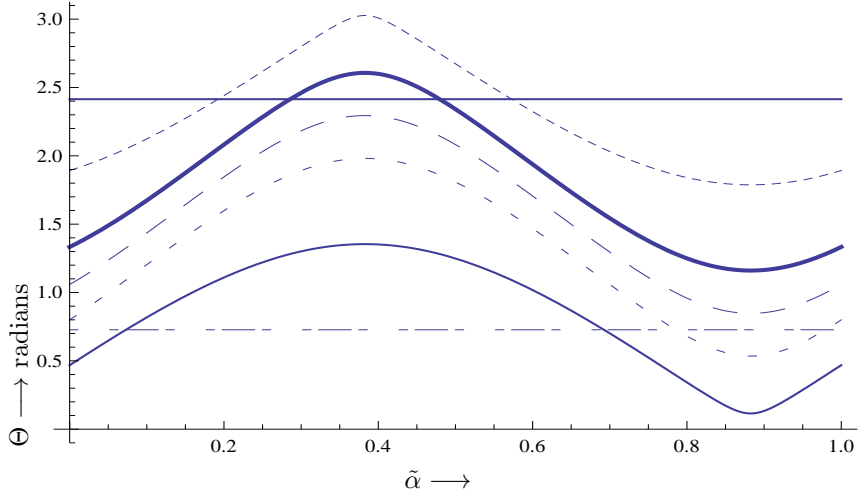}\\
\hspace{-0.0cm} {$\tilde{\alpha}\longrightarrow$}
\caption{ Due to the diurnal motion of the Earth different angles $\Theta$ in galactic coordinates are sampled as the earth rotates. The angle $\Theta$  scanned by the direction of observation is shown for various inclinations $\tilde{\delta}$. 
We see that, for negative inclinations, the angle $\Theta$ can take values near $\pi$, i.e. opposite to the direction of the sun's velocity, where the rate attains its maximum. For an explanation of the curves see Fig. \ref{fig:diurnal}}
 \label{fig:theta}
  \end{center}
  \end{figure}

 The  angle $\Theta$  scanned by the direction of observation is shown, for various inclinations $\tilde{\delta}$, in Fig.~\ref{fig:theta}. We see that for  negative inclinations, the angle $\Theta$ can take values near $\pi$, i.e. opposite to the direction of the sun's velocity, where the rate attains its maximum (see Fig. \ref{fig:theta}).   

\begin{figure}[!ht]
 \begin{center}
{\rotatebox{90}{\hspace{-0.0cm} {$\Gamma/\Gamma_{st}$} (sense known)}}
\includegraphics[width=0.9\textwidth,height=0.5\textwidth]{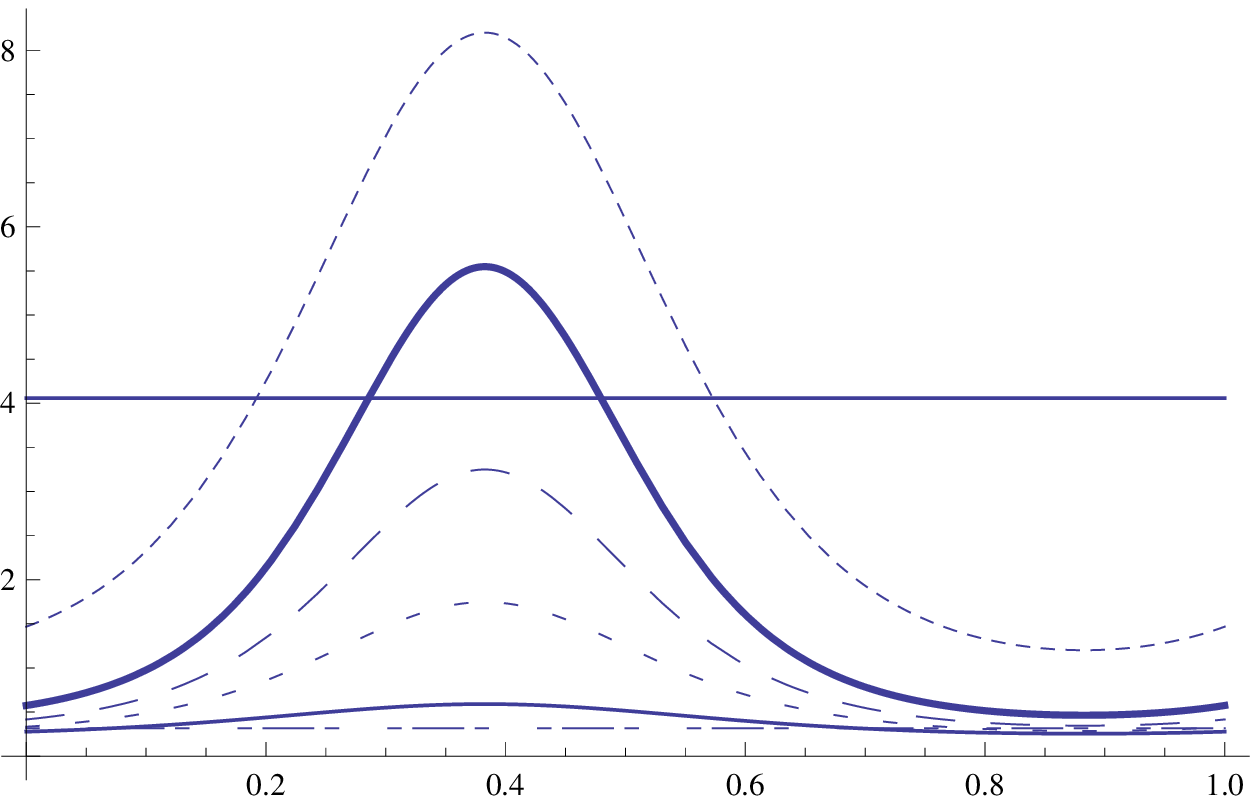}
\hspace{-0.0cm} {$\frac{t}{T} \longrightarrow$}
\\
{\rotatebox{90}{\hspace{-0.0cm} {$\Gamma/\Gamma_{st}$} (sense unknown)}}
\includegraphics[width=0.9\textwidth,height=0.5\textwidth]{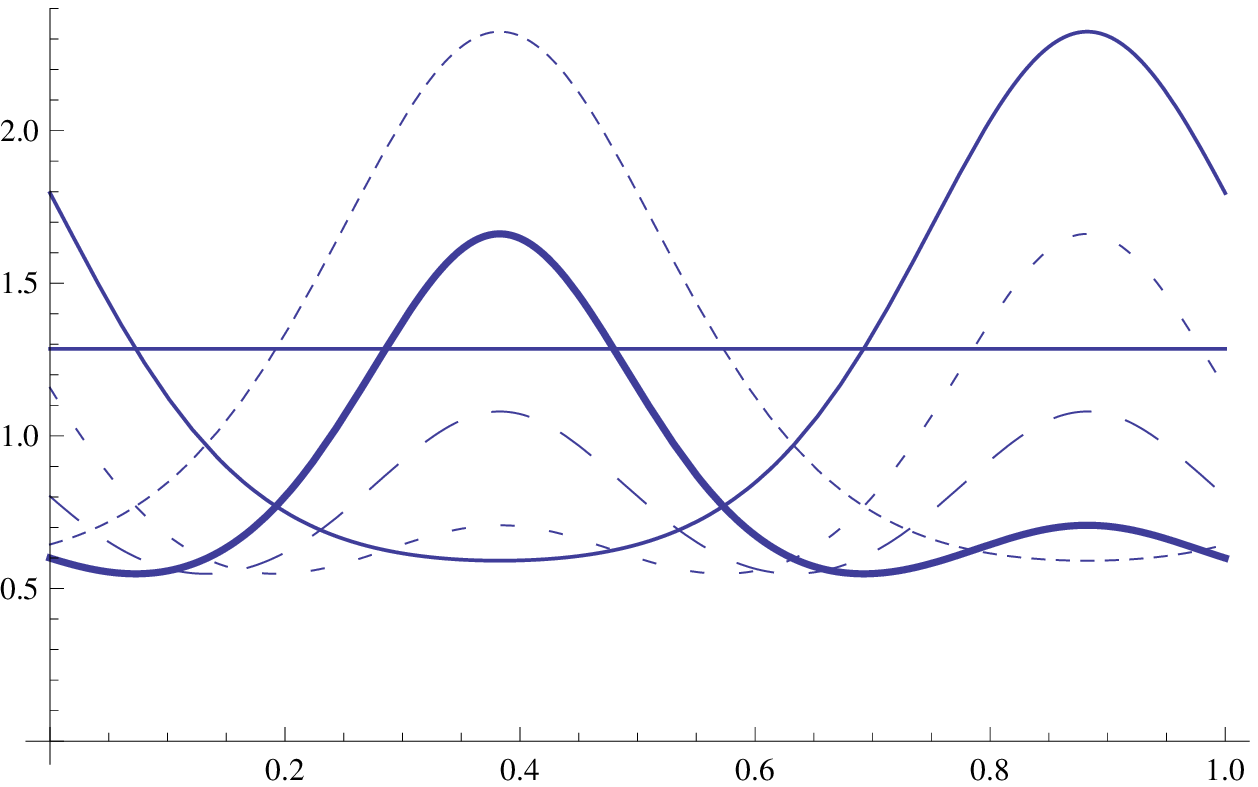}
\hspace{-0.0cm} {$\frac{t}{T} \longrightarrow$}
\\
\caption{The time dependence (in units of the Earth's rotation period)   of the ratio of the directional width divided by the non directional width for various inclinations $\tilde{\delta}$, when the sense can be determined (top) or both senses are included (bottom). In The curves indicated by intermediate thickness solid, the short dash, thick solid line, long dashed, dashed, fine solid line,  and the long-short dashed correspond to inclination $\tilde{\delta}=-\pi/2,-3\pi/10,-\pi/10,0,\pi/10,3\pi/10$ and $\pi/2$  respectively. We see that, for negative inclinations, the angle $\Theta$ can take values near $\pi$, i.e. opposite to the direction of the sun's velocity, where the rate attains its maximum if the sense of direction is known. There is no time variation, of course, when $\tilde{\delta}=\pm\pi/2$. }
 \label{fig:diurnal}
  \end{center}
  \end{figure}

The equipment scans different parts of the galactic sky, i.e. observes different angles $\Theta$. So the rate will change with time depending on whether the sense of of observation. We assume that the sense of direction can be distinguished in the experiment. 
The total flux  is exhibited in Fig. \ref{fig:diurnal}.
	\section{Discussion}
	In the present work we discussed the time variation of the width of of the axion to photon resonance cavities involved in Axion Dark Matter Searches. We find two important signatures:
	\begin{itemize}
	\item Annual variation due to the motion of the Earth around the sun. We find that  in the relative width, i.e. the width divided by its time average, can attain differences of  about $15\%$ between the maximum expected in June  and the minimum  expected six months later. This variation is larger than the modulation expected  in ordinary dark matter of WIMPs. It does not depend on the geometry of the cavity or other details of the apparatus. It does not depend strongly on the assumed velocity distribution.
	\item A characteristic diurnal variation in of the width in directional experiments with most favorable scenario in the range of $m_e L=1.0-2.5\times 10^{-3}$eV\,m. This arises from asymmetries of the local axion velocity with respect to the sun's direction of motion manifested in a time dependent way due to the rotation of the Earth around its own axis. Admittedly such experiments are much  harder, but the expected signature persists, even if one cannot tell the direction of motion of the axion velocity entering in the expression of the width.  Anyway once such a device is operating, data  can be taken  as usual. Only one has to  bin them according the  time they were obtained. If a potentially useful signal is found,  a complete analysis can be done according the directionality to firmly establish that the signal is due to the axion.
	\end{itemize}
	In conclusion in this work we have elaborated on two signatures that might aid the analysis of axion dark matter searches.
	
{\bf	Acknowledgments}:
IBS-Korea partially supported this project under system code IBS-R017-D1-2014-a00.

\end{document}

%% file: begin.tex
\newcommand{\beqa}{\begin{eqnarray}}
\newcommand{\eeqa}{\end{eqnarray}}
\newcommand{\bdm}{\begin{displaymath}}
\newcommand{\edm}{\end{displaymath}}

\renewcommand{\thefootnote}{\#\arabic{footnote}}
\renewcommand{\theequation}{\thesection.\arabic{equation}}
\renewcommand{\thefigure}{\thesection.\arabic{figure}}
\renewcommand{\thetable}{\thesection.\arabic{table}}